\documentstyle{article}

\newcommand{\LL}{{\rm L}}
\newcommand{\MM}{{\rm M}}

\newcommand{\ii}{{\rm i}}

\newcommand{\dd}{{\rm d}}
\newcommand{\ee}{{\rm e}}
\newcommand{\simg}{\stackrel{>}{_\sim}}
\newcommand{\siml}{\stackrel{<}{_\sim}}
\newcommand{\kk}{k}
\newcommand{\RR}{R}
\begin{document}
\begin{center}
{\large\bf Metamagnetism and Fermi Surface \\
in the Anderson Lattice Model} \\
\vspace*{1em}
{Yoshiaki \=ONO} \\
\vspace*{1em}
Department of Physics, Nagoya University,\\
 Furo-cho, Chikusa-ku, Nagoya 464-01, Japan \\
\end{center}
\vspace*{1em}
\begin{center}
{\bf ABSTRACT}\\
\end{center}
\vspace*{1em} \noindent

We investigate magnetization as functions of external magnetic field $H$ in the
$U$-infinite Anderson
lattice model within the leading order approximation in the $1/N$-expansion.
At $T=0$, at $H=H_\MM$ where the Zeeman energy is equal to a certain
characteristic energy in the
system, the magnetization curve has a kink and the differential susceptibility
$\dd M/\dd H$ shows a
jump.
At finite temperature, $\dd M/\dd H$ shows a peak around $H_\MM$. Its maximum
value increases with
decreasing $T$ and saturates to a finite value at $T\to 0$.
When $H<H_\MM$, the $f$ and the conduction electrons form the renormalized
bands with a large Fermi
surface determined by the Luttinger sum rule.
On the other hand, when $H>H_\MM$, the bands reform themselves significantly
free from the Luttinger
sum rule, eventually leading to a small Fermi surface at $H \gg H_\MM$.
The results are consistent with the metamagnetic properties observed in the
heavy fermion
CeRu$_2$Si$_2$.

\clearpage

There have been extensive studies on the metamagnetic behavior observed in the
normal metallic states
of the heavy fermion compounds. The investigations have been mostly focused on
CeRu$_2$Si$_2$,\cite{Haen,Mig,Aoki,Sakaki} whose magnetization shows an abrupt
increase with
increasing $H$ at a certain critical value, $H=H_\MM \sim 7.7$T. Many other
properties of the material
also show pronounced changes at $H\simeq H_\MM$.
The drastic is the change of the Fermi surface observed in the recent de
Haas-van Alphen (dHvA)
experiments:\cite{Aoki} the large Fermi surface observed at $H<H_\MM$ seems to
suggest existence of
the itinerant $f$-electrons, while the small Fermi surface observed at
$H>H_\MM$ seems to suggest the
localized $f$-electrons.
Several theoretical studies have been already done on these
problems.\cite{Brenig,Ohkawa,Konno,Miyake,Evans,Saso} They comment on the
magnetization curves but
nothing on the drastic change of the Fermi surface.

The purpose of the present paper is to investigate theoretically the
metamagnetic properties including
both the magnetization and the Fermi surface. We employ the $U$ - infinite
Anderson lattice model in the
auxiliary boson representation.\cite{Coleman}
Previously, the model has been studied in the limit of zero external magnetic
field within the approximation
including terms of the leading order in the expansion with respect to the
inverse of the spin-orbit
degeneracy ($1/N$-expansion) under the strict local constraints guaranteeing
the equivalence of the
bosonic version to the original $U$-infinite model.\cite{O}
The results very well account for the various properties of the heavy fermion
systems in the high
temperature incoherent regime exhibiting the Kondo effects, the low temperature
coherent regime with
heavy fermions and the intermediate temperature regime showing the crossover
behavior between the
two limiting regimes in a unified way.
The present study is a straightforward extension of the previous study to cases
with finite external
magnetic field.

Our Hamiltonian is given by
\begin{eqnarray}
{\cal H} = \sum_{m=-J}^J \sum_{\kk_m} \epsilon_{\kk_m}c_{\kk_m}^+c_{\kk_m}
  + \sum_i \sum_{m=-J}^J (\epsilon_f-mH) f_{im}^+f_{im} \nonumber \\
   + N_\LL^{-\frac12} \sum_i \sum_{m=-J}^J
    \sum_{\kk_m} (V_{\kk_m} e^{-\ii \kk_m \cdot \RR_i} c_{\kk_m}^+ f_{im}b_i^+
+ h.c.),  \label{MODEL}
\end{eqnarray}
where $m$ stands for the degrees of freedom due to the spin-orbital degeneracy,
$m=-J,-J+1, \cdots
\ J$ \ \ $(2J+1 \equiv N)$, and $i$ stands for the lattice sites, $i=1,2,
\cdots \ N_\LL$. $k_m$
represents the generalized wave vector specifying wave vector as well as
spin-orbital degrees of
freedom. $c_{\kk_m}^+$ is the creation operator for the $c$-electron.  $b_i^+$
and $f_{im}^+$ are the
creation operators for the slave-boson (SB) and the pseudo-fermion (PF)
representing the empty and
the singly occupied states of the $i$-th $f$-site, respectively.
Here we set $g_f\mu_{\rm B}=1$ and $g_c\mu_{\rm B}=0$. Then, $mH$ is the Zeeman
energy for the
$f$-electron under the external magnetic field $H$.\cite{gval} The energies
$\epsilon_{\kk_m}$ and
$\epsilon_f$ are measured relative to the chemical potential $\mu$.

This model (\ref{MODEL}) is equivalent to the original $U$-infinite Anderson
lattice model as long as it is
treated within the physical space where the following local constraints hold:
\begin{equation}  \label{LC}
 \hat{Q}_i \equiv \sum_{m=-J}^J f_{im}^+f_{im} + b_i^+b_i = 1,
\ \ \ \ (i=1,2,\cdots \ N_\LL).
\end{equation}
The expectation value of an operator $\hat{O}$ calculated under the local
constraint (\ref{LC}) is given
by\cite{Coleman}
\begin{equation} \label{AV}
  \langle\hat{O}\rangle = \lim_{\{\lambda_i\}\to\infty}
 \langle\hat{O}\prod_i\hat{Q}_i\rangle_\lambda/
 \langle\prod_i\hat{Q}_i\rangle_\lambda,
\end{equation}
where
$  \langle\hat{A}\rangle_\lambda\equiv
  \mbox{Tr}[e^{-\beta \cal H_{\lambda}}\hat{A}]/
  \mbox{Tr}[e^{-\beta \cal H_{\lambda}}] $ \
with
${\cal H}_{\lambda}\equiv{\cal H}+\sum_i \lambda_i \hat{Q}_i$.
For calculating $\langle\hat{A}\rangle_\lambda$, we treat $V_{\kk_m}$ as a
perturbation and employ the
standard perturbation method using the Feynman diagrams together with the
expansion ($1/N$-expansion)
from the large limit of the spin-orbit degeneracy $N$, while keeping the total
degrees of freedom for the
$c$-electrons to be constant:
$N_\LL^{-1}\sum_m \sum_{\kk_m} 1 = 2$.
Following the procedure mentioned above we calculate the single particle
Green's functions within the
leading order in power of $1/N$. The explicit forms of the $c$-electron, the SB
and the PF Green's
functions are given, respectively, by\cite{O}
\begin{eqnarray}
G_{\kk_m}(\ii\omega_n) &=&
[\ii\omega_n-\epsilon_{\kk_m}-\Sigma_{\kk_m}(\ii\omega_n)]^{-1},
\label{G} \\
B(\ii\nu_n-\lambda_i) &=& [\ii\nu_n-\lambda_i-\sum_m
\Pi_m(\ii\nu_n-\lambda_i)]^{-1}, \label{B} \\
F_m(\ii\omega_n-\lambda_i) &=& [\ii\omega_n-\lambda_i-\epsilon_f+mH]^{-1},
\label{F}
\end{eqnarray}
where the self-energy parts are given by
\begin{eqnarray}
\lefteqn{\Sigma_{\kk_m}(\ii\omega_n)
= |V_{\kk_m}|^2 \lim_{\{\lambda_i\}\to\infty}
       [-T}       \nonumber \\
&\times&
\sum_{\nu_{n'}}F_m(\ii\omega_n+\ii\nu_{n'}-\lambda_i)B(\ii\nu_{n'}-\lambda_i)/\langle\hat{Q}_i\rangle_
\lambda], \label{Sigma} \\
\lefteqn{\Pi_m(\ii\nu_n-\lambda_i)
= N_\LL^{-1} } \nonumber \\
&\times&  \sum_{\kk_m}|V_{\kk_m}|^2
T\sum_{\omega_{n'}}F_m(\ii\omega_{n'}+\ii\nu_n-\lambda_i)G_{\kk_m}(\ii\omega_{n'}), \label{Pi}
\end{eqnarray}
with
$\langle\hat{Q}_i\rangle_\lambda \equiv \langle b_i^+b_i\rangle_\lambda+\sum_m
\langle
f_{im}^+f_{im}\rangle_\lambda$.
The occupation numbers of the SB and the PF are given in the leading order of
the $1/N$-expansion by
\begin{eqnarray}
\langle b_i^+b_i\rangle_\lambda
   &=& -T\sum_{\nu_n}B(\ii\nu_n-\lambda_i), \label{bb} \\
\langle f_{im}^+f_{im}\rangle_\lambda
   &=& T\sum_{\omega_n}F_m(\ii\omega_n-\lambda_i) \nonumber \\
   &+& T\sum_{\nu_n}B(\ii\nu_n-\lambda_i)
       \frac{\dd \Pi_m(\ii\nu_n-\lambda_i)}{\dd (\ii\nu_n)}.
   \label{ff}
\end{eqnarray}
Note that $F_m$ is not modified in the lowest order of $1/N$ as shown in
eq.(\ref{F}), while the
correction of $O(1/N)$ to $\langle f_{im}^+f_{im}\rangle_\lambda$ must be
included in eq.(\ref{ff}) to
obtain $\langle\hat{Q}_i\rangle_\lambda$ of $O((1/N)^0)$.
Equations (\ref{G})-(\ref{ff}) constitute a set of self-consistent equations
(SCE), which was already
solved in the limit of $H=0$ analytically at $T=0$ and numerically at finite
temperatures.\cite{O}
Now, we solve the SCE with $H\neq0$.

To begin with, we investigate the magnetization process at $T=0$. With
$H\neq0$, we rewrite the
occupation numbers, eqs.(\ref{bb}) and (\ref{ff}), and the sum of them,
$\langle\hat{Q}_i\rangle_\lambda$, in the low temperature limit as
\begin{eqnarray}
\langle b_i^+b_i\rangle_\lambda
   &\stackrel{\lambda_i\to\infty}{\longrightarrow}&
   a\ee^{-\beta(\lambda_i+\epsilon_f-E_0)}, \nonumber \\
\langle f_{im}^+f_{im}\rangle_\lambda
   &\stackrel{\lambda_i\to\infty}{\longrightarrow}&
   \ee^{-\beta(\lambda_i+\epsilon_f-mH)}
   + \Delta n_{fm}\ee^{-\beta(\lambda_i+\epsilon_f-E_0)},
   \nonumber \\
   \langle \hat{Q}_i\rangle_\lambda
   &\stackrel{\lambda_i\to\infty}{\longrightarrow}&
   \sum_m \ee^{-\beta(\lambda_i+\epsilon_f-mH)}
   + \ee^{-\beta(\lambda_i+\epsilon_f-E_0)},
   \nonumber \\    \label{Q}
\end{eqnarray}
with
$\Delta n_{fm} \equiv -a\left.\frac{\rm d}{{\rm
d}\omega}\mbox{Re}\Pi_m(\omega)\right|_{\omega=\epsilon_f-E_0}$,
where $E_0$ and $a$ are, respectively, the binding energy and the residue of
the resonance state in the
SB spectrum, Im$B(\nu-\lambda_i+i0^+)$, which are determined by the relations:
\begin{eqnarray}
  \epsilon_f-E_0
         -\sum_m{\rm Re}\Pi_m(\epsilon_f-E_0)=0 \label{E0}, \\
\frac1a = 1-\left.\sum_m\frac{\rm d}{{\rm d}\omega}
    {\rm Re}\Pi_m(\omega)\right|_{\omega=\epsilon_f-E_0}. \label{AR}
\end{eqnarray}
In the above, both $E_0$ and $a$ are functions of $T$ and $H$ to be determined
later. $E_0^0\equiv
E_0|_{T=H=0}$ represents the characteristic energy of the system, which
corresponds to the Kondo
temperature defined in the single impurity Anderson model.\cite{O}
We note that the continuum in the SB spectrum, which has finite intensity for
$\nu>\lambda_i+\epsilon_f-min(E_0,JH)$, is irrelevant to calculate the
occupation numbers at $T=0$.
{}From eq.(\ref{AV}), the average number of the $f$-electron with $m$-th orbit
is given by
$n_{fm}=\lim_{\lambda_i\to\infty}[\langle f_{im}^+f_{im}\rangle_\lambda/\langle
\hat{Q}_i\rangle_\lambda]$.
Using eq.(\ref{Q}), we obtain at $T=0$ as
\begin{eqnarray}
n_{fm}=\frac{\Delta n_{fm}}{1+\alpha}+n_{fJ}^0\delta_{m,J},
       \label{nfm}
\end{eqnarray}
where
$n_{fJ}^0=\alpha/(1+\alpha)$
is the incoherent part of the $f$-electron with $m=J$, which is due to the
zeroth order term w.r.t.
$V_{\kk_m}$, and
$\alpha \equiv \lim_{T\to 0}\ee^{\beta(JH-E_0)}$.
Equation (\ref{nfm}) yields the average number, $n_f=\sum_m n_{fm}$, and the
magnetization,
$M=\sum_m m n_{fm}$, of the $f$-electron, respectively, at $T=0$:
\begin{eqnarray}
n_f&=&(1-a)/(1+\alpha)+n_{fJ}^0, \label{nf} \\
M&=&\sum_m m\Delta n_{fm}/(1+\alpha) +Jn_{fJ}^0. \label{M}
\end{eqnarray}
Substituting eqs.(\ref{B}), (\ref{F}) and (\ref{Q}) into eq.(\ref{Sigma}), we
find
\begin{eqnarray}
\Sigma_{\kk_m}(\ii\omega_n)
&=&  \frac{a|V_{\kk_m}|^2}{\ii\omega_n-E_{0m}}
     \frac{1}{1+\alpha} \nonumber \\
&-&  |V_{\kk_J}|^2
     B(\epsilon_f-JH-\ii\omega_n) n_{fJ}^0\delta_{m,J},
     \label{Sik}
\end{eqnarray}
at $T=0$, with $E_{0m}\equiv E_0-mH$. Substituting eq.(\ref{Sik}) to
eq.(\ref{G}), we obtain the
$c$-electron Green's function.

Now we solve eqs.(\ref{E0}) and (\ref{AR}) with eq.(\ref{Pi}) at $T=0$ to
obtain $E_0$ and $a$ together
with $\alpha$ for given $H$.  Hereafter for the numerical calculations we
assume simple
$k_m$-dependence as:
$\epsilon_{\kk_m}=\epsilon_c-\xi_{\kk_m}$ and $|V_{\kk_m}|^2=V^2\xi_{\kk_m}$
together with a square
density of states:
$\rho^0_m(\xi)\equiv N_\LL^{-1}\sum_{\kk_m}\delta(\xi-\xi_{\kk_m})=2/N$ for
$0\le\xi\le 1$, and
$\rho^0_m(\xi)=0$ for otherwise.
The parameters are chosen as: $J=\frac52 \ (N=6), n=n_c+n_f=1.5,
\epsilon_c-\epsilon_f=0.9$ and
$V=0.2$.\cite{Parm}

In Fig.1, we show the self-consistent solutions at $T=0$ for $a$ and $\alpha$
as functions of $H$, and
those for $E_0$ by the dashed line in Fig.2.
We find that, at $H<H_\MM$, $E_0>JH$ and $\alpha=0$, while, at $H>H_\MM$,
$E_0=JH$ and
$\alpha\ne0$, where $H_\MM=11.12\times 10^{-4}$ is of order of $E_0^0/2$.
($E_0^0=22.77\times
10^{-4}$ as plotted in Fig.2.)
Each of $a$, $\alpha$ and $E_0$ has a kink at $H=H_\MM$. Correspondingly, the
magnetization curve
has a kink at $H=H_\MM$ as shown in Fig.3 (see the dashed line for $T=0$). At
$H=H_\MM$, the slope
of $M$ increases abruptly with increasing $H$, because so does that of the
incoherent part $n_{fJ}^0$
(or $\alpha$) as shown in Fig.1 contributing to the 2nd term in eq.(\ref{M}).

Next, we investigate the magnetization process at finite temperature. In this
case, we can not use the
analytic expressions eqs.(\ref{nfm})-(\ref{Sik}) obtained at $T=0$. Then we
solve the SCE
(\ref{G})-(\ref{ff}) numerically in the same way as taken in the previous
studies.\cite{O} Fig.2 shows
$H$-dependence of $E_0$ calculated from eq.(\ref{E0}) at several temperatures.
When $H>H_\MM$, we
observe that $(E_0-JH)\propto T$ at the low temperatures, and find that the
extrapolated value of
$\ee^{\beta(JH-E_0)}$ at $T\to 0$ coincides with $\alpha$ calculated through
the SCE at $T=0$ (see
also Fig.1).
In Fig.3, the magnetization curves are plotted at several temperatures. At the
lower temperatures, the
slope of $M$ increases significantly around $H=H_\MM$, while, at the higher
temperatures, $M$
increases only monotonically. In Fig.4, we plot the differential susceptibility
defined by $\dd M/\dd H$ as
functions of $H$ at several temperatures.  At the lower temperature $T\siml
5\times 10^{-4}$, $\dd
M/\dd H$ shows a peak around $H=H_\MM$. Its maximum value increases with
decreasing $T$ and
saturates to a finite value at $T\to 0$.
Such behaviour in $\dd M/\dd H$ at $T\to 0$ has been observed in the recent
magnetization
measurements at very low temperature.\cite{Sakaki}

The magnetic susceptibility $\chi_s$ is plotted as function of $T$ in Fig.5,
where
$\chi_s^0=J(J+1)n_f^0/3T$ is the contribution due to the incoherent part of the
$f$-electron,
$n_f^0=\sum_m n_{fm}^0$, while $\Delta\chi_s$ is the coherent part due to the
lowest order corrections
in the $1/N$-expansion.
We observe a maximum in $\chi_s$ at $T=T_{\max}$, where
$T_{\max}\sim5.5\times10^{-4}$ is of order
of $E_0^0/4$.
At $T\ll T_{\max}$, $\Delta\chi_s$ is the dominant term leading to the enhanced
Pauli paramagnetism,
$\Delta\chi_s \sim 1/E_0^0$, while, at $T\gg T_{\max}$, $\chi_s^0$ is the
dominant term leading to the
Curie law, $\chi_s^0\sim J(J+1)/3T$.\cite{O}
At $T\simg T_{\max}$, the incoherent part, $n_f^0$, contributes dominantly to
$M$ even for $H=0$. Thus,
the metamagnetic behaviour due to the rapid increase of $n_{fJ}^0$ is no more
distinguished at $T\simg
T_{\max}$, as seen in Fig.4 and as observed in the experiments.\cite{Haen}
We emphasize that both $H_\MM$ and $T_{\max}$ are proportional to $E_0^0$ which
determine the Fermi
liquid properties of the system, {\it e.g.}, $\chi_s|_{T=0}\sim1/E_0^0$. Such
relationship among $H_\MM$,
$T_{\max}$ and the characteristic energy of the Fermi liquid has been observed
in the pressure
dependence of metamagnetic properties.\cite{Mig}

Finally, we discuss the quasi-particle properties and the Fermi surface.
Substituting eq.(\ref{Sik}) into eq.(\ref{G}), we obtain the $c$-electron
Green's function  at $T=0$ for
$H<H_\MM$ as
$G_{\kk_m}(i\omega_n)=\sum_{\gamma=\pm}A_{\kk_m}^{\gamma}[\ii\omega_n-E_{\kk_m}^{\gamma}]^{-1
}$,
with
\begin{eqnarray}
E_{\kk_m}^{\pm} &\equiv&
  \frac12[\epsilon_{\kk_m}+E_{0m}\pm
  \sqrt{(\epsilon_{\kk_m}-E_{0m})^2+4a|V_{\kk_m}|^2}],
  \nonumber \\
A_{\kk_m}^{\pm} &\equiv&
  (E_{\kk_m}^{\pm}-E_{0m})/
  (E_{\kk_m}^{\pm}-E_{\kk_m}^{\mp}). \nonumber
\end{eqnarray}
In Fig.6(a), $E_{\kk_m}^{\pm}$ are plotted as functions of $\xi_{\kk_m}$ for
$H<H_\MM$, which describe
the coherently $cf$-hybridized excitation with the Zeeman split resonance level
$E_{0m}$.
On the other hand, for $H>H_\MM$, the imaginary part of the $c$-electron
self-energy with $m=J$ is
finite at $\omega<0$ as shown in eq.(\ref{Sik}) (note that the continuum in the
SB spectra is finite for
$\nu>\lambda_i+\epsilon_f-min(E_0,JH)$). Therefore the $cf$-hybridized
excitation with $m=J$ becomes
incoherent. However the incoherent excitation has a finite energy gap as seen
in Fig.6(b), because the
resonance level $E_{0J}=0$. Thus they are irrelevant to the low energy
properties at $T=0$.

Using eqs.(\ref{Sigma}) and (\ref{ff}), the average number of the $f$-electron
with $m$-th orbit is
written by:
\begin{eqnarray}
n_{fm}= \frac1{N_\LL}\sum_{\kk_m}
   \left[-T\sum_{\omega_n}G_{\kk_m}(\ii\omega_n)
   \frac{\dd\Sigma_{\kk_m}(\ii\omega_n)}{\dd(\ii\omega_n)}\right]
   +n_{fm}^0. \nonumber
\end{eqnarray}
Then the total number of the $c$- and the $f$-electrons per site is given by:
\begin{eqnarray}
n_c+n_f = \frac1{N_\LL}
   \sum_{m,\kk_m}\left[T\sum_{\omega_n}
   \frac{\dd\log\{G_{\kk_m}(\ii\omega_n)^{-1}\}}
   {\dd(\ii\omega_n)}\right] +n_f^0. \nonumber
\end{eqnarray}
When $T=0$ and $H<H_\MM$, it yields
$n_c+n_f = N_\LL^{-1} \sum_m\sum_{\kk_m}\theta(-E_{\kk_m}^-)$,
by using $n_f^0=0$ and ${\rm Im}\Sigma_{\kk_m}(\ii0_+)=0$.
Thus the Fermi surface is determined by the total number of the $c$- and the
$f$-electrons, leading to
a large Fermi surface as seen in Fig.6(a). This is nothing but the Luttinger
sum rule.\cite{Lut}
On the other hand, when $T=0$ and $H>H_\MM$, $n_f^0=n_{fJ}^0\ne 0$. Therefore,
the Luttinger sum
rule breaks down.
As $n_{fJ}^0$ increases with increasing $H$, the phase volume enclosed by the
Fermi surface with $m\ne
J$ decreases abruptly, leading to a small Fermi surface determined almost only
by the $c$-electrons as
seen in Fig.6(b).
The drastic change of the Fermi surface around $H=H_\MM$ has been observed in
the recent dHvA
experiments.\cite{Aoki}

In conclusion, we have investigated the magnetization process of the Anderson
lattice model in the
metallic phase within the leading order approximation in the $1/N$-expansion.
A critical field $H_\MM$ has been shown to exist at $T=0$: at $H>H_\MM$, the
incoherent part of the
$f$-electron due to the zeroth order term w.r.t. the $cf$-mixing becomes
relevant to the physical
quantities, although it is irrelevant at $H<H_\MM$.
Consequently, the magnetization increases far more steeply with increasing $H$
for $H>H_\MM$, and
the differential susceptibility shows a jump at $H=H_\MM$.
Moreover the phase volume enclosed by the Fermi surface has been found to
change drastically at
$H=H_\MM$: at $H<H_\MM$, it is as large as determined by the Luttinger sum
rule, while, at
$H>H_\MM$, it shrinks leading to a small Fermi surface determined almost only
by the $c$-electrons.
At finite temperature, the smooth metamagnetic behavior has been observed at
$T\siml T_{\max}$, the
temperature where the magnetic susceptibility has a maximum.
The results agree with the observation in the metamagnetic heavy fermion
CeRu$_2$Si$_2$.

The author would like to thank Y. Kuroda, T. Matsuura, Y. \=Onuki and T.
Sakakibara for useful discussions
and comments. This work was partly supported by the Grant-in-Aid for Scientific
Research from the
Ministry of Education, Science and Culture.

\noindent
\begin{flushleft}
{\Large\bf Figure Captions}
\end{flushleft}
\begin{description}

\item[Fig.1]
Self-consistent solutions for the residue $a$ of the resonance level of the
slave-boson (dotted line) and
$\alpha$ defined in eq.(\ref{nfm}) (solid line) at $T=0$ as functions of $H$.

\item[Fig.2]
Self-consistent solutions for the binding energy $E_0$ of the resonance level
of the slave-boson at
$T=0$ (dotted line) and at several finite temperatures (solid lines) as
functions of $H$.
\item[Fig.3]
Magnetization $M$ at $T=0$ (dotted line) and at several finite temperatures
(solid lines) as functions of
$H$.

\item[Fig.4]
Differential susceptibility $\dd M/\dd H$ at $T=0$ (dotted line) and at several
finite temperatures (solid
lines) as functions of $H$.

\item[Fig.5]
Temperature dependence of the magnetic susceptibility. The dashed line is the
contribution of the
incoherent part of the $f$-electron, $\chi_s^0$, the dotted line is that of the
coherent part of
$O((1/N)^0)$, $\Delta\chi_s$, and the solid line is the sum of them, $\chi_s$.

\item[Fig.6]
Schematic structures of the renormalized bands $E_k^\pm$ for $m=-J,-J+1, \cdots
\ J$ at
$H<H_\MM$ (a) and $H>H_\MM$ (b) as functions of $\xi_k$. The dashed lines in
(b) are for $m=J$.
The bare $c$-electron band $\epsilon_k$ is also plotted (dotted line).

\end{description}


\begin{thebibliography}{99}
\bibitem{gval}
If we assume $g_c=g_f=g$, we obtain almost the same results for the present
case, because the
magnetization for the $c$-electron $M_c$ is much smaller than that for the
$f$-electron $M_f$:
$M_c/M_f\sim O(E_0^2/aV^2)\ll 1$.\cite{O}
\bibitem{Parm}
We also examine the other sets of parameters and obtain the same results in the
metallic case by
rescaling the energies in unit of $E_0^0$ calculated for each parameter set.
\end{thebibliography}
\end{document}